\documentclass[twocolumn,twocolappendix]{aastex631}

\def\Fig{\mbox{Figure~}}

\def\Tab{\mbox{Table~}}

\def\Sec{\mbox{Section~}}

\shorttitle{Not-so-little red dots}
\shortauthors{Gentile et al.}

\begin{document}

\title{Not-so-little Red Dots: Two massive and dusty starbursts at $z\sim5-7$ pushing the limits of star formation discovered by JWST in the COSMOS-Web survey}

\correspondingauthor{Fabrizio Gentile}
\email{fabrizio.gentile3@unibo.it}

\author[0000-0002-8008-9871]{Fabrizio Gentile}
\affiliation{University of Bologna, Department of Physics and Astronomy (DIFA), Via Gobetti 93/2, I-40129, Bologna, Italy}
\affiliation{INAF -- Osservatorio di Astrofisica e Scienza dello Spazio, via Gobetti 93/3 - 40129, Bologna - Italy}

\author[0000-0002-0930-6466]{Caitlin M. Casey}
\affiliation{The University of Texas at Austin, 2515 Speedway Boulevard Stop C1400, Austin, TX 78712, USA}


\author[0000-0003-3596-8794]{Hollis B. Akins}
\affiliation{The University of Texas at Austin, 2515 Speedway Boulevard Stop C1400, Austin, TX 78712, USA}

\author[0000-0002-3560-8599]{Maximilien Franco}
\affiliation{The University of Texas at Austin, 2515 Speedway Boulevard Stop C1400, Austin, TX 78712, USA}

\author[0000-0002-6149-8178]{Jed McKinney}
\affiliation{The University of Texas at Austin, 2515 Speedway Boulevard Stop C1400, Austin, TX 78712, USA}


\author[0000-0002-8494-3123]{Edward Berman}
\affiliation{Department of Physics, Northeastern University, 360 Huntington Ave, Boston, MA}

\author[0000-0003-3881-1397]{Olivia R. Cooper}\altaffiliation{NSF Graduate Research Fellow}
\affiliation{The University of Texas at Austin, 2515 Speedway Boulevard Stop C1400, Austin, TX 78712, USA}

\author[0000-0003-4761-2197]{Nicole E. Drakos}
\affiliation{Department of Physics and Astronomy, University of Hawaii, Hilo, 200 W Kawili St, Hilo, HI 96720, USA}

\author[0000-0002-3301-3321]{Michaela Hirschmann}
\affiliation{Institute of Physics, GalSpec, Ecole Polytechnique Federale de Lausanne, Observatoire de Sauverny, Chemin Pegasi 51, 1290 Versoix, Switzerland}
\affiliation{INAF, Astronomical Observatory of Trieste, Via Tiepolo 11, 34131 Trieste, Italy}

\author[0000-0002-7530-8857]{Arianna S. Long}\altaffiliation{NASA Hubble Fellow}
\affiliation{The University of Texas at Austin, 2515 Speedway Boulevard Stop C1400, Austin, TX 78712, USA}

\author[0000-0002-4872-2294]{Georgios Magdis}
\affiliation{Cosmic Dawn Center (DAWN), Denmark} 
\affiliation{DTU-Space, Technical University of Denmark, Elektrovej 327, 2800, Kgs. Lyngby, Denmark}
\affiliation{Niels Bohr Institute, University of Copenhagen, Jagtvej 128, DK-2200, Copenhagen, Denmark}

\author[0000-0002-6610-2048]{Anton M. Koekemoer}
\affiliation{Space Telescope Science Institute, 3700 San Martin Dr., Baltimore, MD 21218, USA} 

\author[0000-0002-5588-9156]{Vasily Kokorev}
\affiliation{Kapteyn Astronomical Institute, University of Groningen, PO Box 800, 9700 AV Groningen, The Netherlands}

\author[0000-0002-7087-0701]{Marko Shuntov}
\affiliation{Cosmic Dawn Center (DAWN), Denmark} 
\affiliation{Niels Bohr Institute, University of Copenhagen, Jagtvej 128, DK-2200, Copenhagen, Denmark}

\author[0000-0003-4352-2063]{Margherita Talia}
\affiliation{University of Bologna, Department of Physics and Astronomy (DIFA), Via Gobetti 93/2, I-40129, Bologna, Italy}
\affiliation{INAF -- Osservatorio di Astrofisica e Scienza dello Spazio, via Gobetti 93/3 - 40129, Bologna - Italy}


\author[0000-0001-9610-7950]{Natalie Allen}
\affiliation{Cosmic Dawn Center (DAWN), Denmark} 
\affiliation{Niels Bohr Institute, University of Copenhagen, Jagtvej 128, DK-2200, Copenhagen, Denmark}

\author[0000-0003-0129-2079]{Santosh Harish}
\affiliation{Laboratory for Multiwavelength Astrophysics, School of Physics and Astronomy, Rochester Institute of Technology, 84 Lomb Memorial Drive, Rochester, NY 14623, USA}

\author[0000-0002-7303-4397]{Olivier Ilbert}
\affiliation{Aix Marseille Univ, CNRS, CNES, LAM, Marseille, France  }

\author[0000-0002-9489-7765]{Henry Joy McCracken}
\affiliation{Institut d’Astrophysique de Paris, UMR 7095, CNRS, and Sorbonne Université, 98 bis boulevard Arago, F-75014 Paris, France}

\author[0000-0001-9187-3605]{Jeyhan S. Kartaltepe}
\affiliation{Laboratory for Multiwavelength Astrophysics, School of Physics and Astronomy, Rochester Institute of Technology, 84 Lomb Memorial Drive, Rochester, NY 14623, USA}

\author[0000-0001-9773-7479]{Daizhong Liu}
\affiliation{Purple Mountain Observatory, Chinese Academy of Sciences, 10 Yuanhua Road, Nanjing 210023, China}

\author[0000-0003-2397-0360]{Louise Paquereau} 
\affiliation{Institut d’Astrophysique de Paris, UMR 7095, CNRS, and Sorbonne Université, 98 bis boulevard Arago, F-75014 Paris, France}

\author[0000-0002-4485-8549]{Jason Rhodes}
\affiliation{Jet Propulsion Laboratory, California Institute of Technology, 4800 Oak Grove Drive, Pasadena, CA 91001, USA}

\author[0000-0003-0427-8387]{R. Michael Rich}
\affiliation{Department of Physics and Astronomy, UCLA, PAB 430 Portola Plaza, Box 951547, Los Angeles, CA 90095-1547}

\author[0000-0002-4271-0364]{Brant E. Robertson}
\affiliation{Department of Astronomy and Astrophysics, University of California, Santa Cruz, 1156 High Street, Santa Cruz, CA 95064, USA}

\author[0000-0003-3631-7176]{Sune Toft}
\affiliation{Cosmic Dawn Center (DAWN), Denmark} 
\affiliation{Niels Bohr Institute, University of Copenhagen, Jagtvej 128, DK-2200, Copenhagen, Denmark}

\author[0000-0002-0236-919X]{Ghassem Gozaliasl}
\affiliation{Department of Computer Science, Aalto University, PO Box 15400, Espoo, FI-00 076, Finland}
\affiliation{Department of Physics, P.O. Box 64, 00014 University of Helsinki, Finland}

\begin{abstract}
\noindent
 We present the properties of two {candidate} massive ($M_\star\sim10^{11}M_\odot$) and dusty ($A_{\rm v}>2.5$ mag) galaxies at $z=5-7$ in the first 0.28 deg$^2$ of the COSMOS-Web survey. One object is spectroscopically confirmed at $z_{\rm spec}=5.051$, while the other has a robust $z_{\rm phot}=6.7\pm0.3$. Thanks to their extremely red colors ($F277W-F444W\sim1.7$ mag), these galaxies satisfy the nominal color-selection for the widely-studied ``little red dot" (LRD) population with the exception of their spatially-resolved morphologies. The morphology of our targets allows us to conclude that their red continuum is dominated by highly obscured stellar emission and not by reddened nuclear activity. Using a variety of SED-fitting tools {and star formation histories}, we estimate the stellar masses to be {$\log(M_\star)=11.32^{+0.07}_{-0.15}$ $M_\odot$ and $\log(M_\star)=11.2^{+0.1}_{-0.2}$ $M_\odot$, respectively, with a red continuum emission dominated by a recent episode of star formation.} We then compare their number density to the halo mass function to infer stellar baryon fractions of {$\epsilon_\star\sim0.25$} and $\epsilon_\star\sim0.5$. Both are significantly higher than what is commonly observed in lower-\textit{z} galaxies or more dust-obscured galaxies at similar redshifts. With very bright ultra-high-\textit{z} Lyman-Break Galaxies and some non-AGN dominated LRDs, such ``extended" LRDs represent another population that may require very efficient star formation at early times.

\end{abstract}


\section{Introduction}

Since its launch, the James Webb Space Telescope (JWST) has dramatically changed our perspective on the high-\textit{z} Universe. The discovery of a significant population of highly red and compact sources at $z\gtrsim5$ (“little red dots”; LRDs henceforth) is puzzling in their origin. {These objects have been found in most of the first JWST surveys through their color in the long-wavelength (LW) filters of NIRCam (mainly F277W and F444W; see e.g. \citealt{Labbe_23a,Labbe_23b,Akins_23,Matthee_23,Kokorev_24,PerezGonzalez_24}). These analyses have shown that LRDs are predominately at $z\gtrsim5$ {(with few lower-\textit{z} analogs; see e.g. the low fraction of Extremely Red Objects at $z<5$ reported by \citealt{Barro_23})}, are spatially compact, and some have a  seemingly decoupled blue SED rising into the rest-UV {(e.g. \citealt{Labbe_23a,Labbe_23b,Kokorev_24,PerezGonzalez_24})}. Moreover, the first spectroscopic follow-ups showed a significant presence of broad emission lines in their spectra, suggesting the presence of Active Galactic Nuclei (AGN) in their cores with a high $M_{\rm BH}/M_\star$ ratio \citep{Kocevski_23,Furtak_23,Greene_23,Matthee_23}.

Still unclear is the main physical process responsible for the red continuum: is it highly-obscured stellar emission (e.g. \citealt{Labbe_23a,Akins_23,Xiao_23,PerezGonzalez_24,Williams_23}) or reddened radiation emitted by the AGN (e.g. \citealt{Labbe_23b,Greene_23,Kokorev_24})? 

{Modeling the spectral energy distribution (SED) of these galaxies with stellar-only templates, one can obtain near ``Universe-breaking" stellar masses (up to $\sim10^{11}$ M$_\odot$ at $z\sim8$; see e.g. \citealt{Labbe_23a}). These values were in tension with the current cosmological model \citep[e.g.][]{Boylan_23}, unless (after some downward revisions) we assume that the star formation in the high-\textit{z} Universe is much more efficient than what is commonly observed at lower-\textit{z} (see e.g. \citealt{Xiao_23}). Of course, assuming that the reddened emission comes from an AGN accretion disk lowers the stellar mass estimates, but a new problem arises when the number density of these LRDs is compared with the quasar luminosity function \citep[e.g.][]{Shen_20}, as LRDs are 100-1000 times more common, implying an overabundant population of massive black holes at early times \citep{Greene_23}.

One of the main reasons for the uncertainty about the dominant continuum emission in LRDs is their compactness, not allowing us to completely disentangle the nuclear and stellar contribution. For this reason, in this letter, we set out to find sources that fulfill the color selection of LRDs but have unambiguously extended morphologies with no embedded point source emission. By virtue of their spatial extent, we can attribute their continuum to stellar origin and -- therefore -- obtain AGN-unbiased estimation of the stellar masses}. In this letter, we report the discovery of {two {candidate} massive ($M\star\sim10^{11}$ M$_\odot$) galaxies} (one of which already has a spectroscopic redshift from the literature) in the COSMOS-Web survey \citep{Casey_23}. \Sec\ref{sec:data} presents the parent sample from which we select our resolved LRD candidates and the photometry available for them in COSMOS-Web. \Sec\ref{sec:sed_fitting} describes the SED fitting procedure followed to estimate the photo-\textit{z} and physical properties, including stellar masses, of our sources and \Sec\ref{sec:morphology} presents a morphological analysis of our sources. \Sec\ref{sec:discussion} discusses our results, by placing our galaxies in a cosmological context. Finally, we conclude in \Sec\ref{sec:conclusions}. Throughout this study, we assume a flat $\Lambda$CDM cosmology with the parameters reported in \citet{Planck_20}, a \citet{Chabrier_03} Initial Mass Function (IMF), and the AB photometric system \citep{Oke_83}.

\section{Data}
\label{sec:data}
\subsection{JWST Photometry from COSMOS-Web}
\label{sec:JWST_phot}
The NIRCam and MIRI photometry for our sources comes from the COSMOS-Web survey (GO \#1727, PIs Kartaltepe \& Casey; \citealt{Casey_23}), a Cycle 1 treasury program consisting of the NIRCam and MIRI imaging of the central region of the COSMOS field \citep[see e.g.][]{Scoville_07,Koekemoer_07}. A full description of the survey design and main scientific goals can be found in \citet{Casey_23}, while a complete discussion about the data reduction will be included in Franco et al. (in prep.). In this paper, we focus on the first data acquired in the COSMOS-Web survey during JWST Cycle 1 in  January and April 2023, covering $\sim0.28$ deg$^2$ with NIRCam (i.e., 52\% of the total area of the survey).

{The space-based photometry in COSMOS is extracted with \texttt{sourceXtractor++} (\texttt{SE++} in the following; \citealt{Bertin_20,Kummel_20}), a model-based tool for extracting photometry from datasets with different spatial resolution. Detection is
conducted on a $\chi^2$-image \citep{Szalay_99} generated by co-adding the four NIRCam filter maps. A double Sérsic profile \citep{Sersic_63} is fitted to the four NIRCam filter images simultaneously.} Further details about the COSMOS-Web catalogs will be included in Shuntov, Paquereau et al., (in prep.).

By default, \texttt{SE++} photometric uncertainties do not include Poisson noise from the background (see e.g. the discussion in \citealt{Akins_23} and \citealt{Casey_23b}). Therefore, we measure the background noise in empty circular apertures and add it in quadrature to the uncertainties estimated by \texttt{SE++}, following \citet{Casey_23b}.

\subsection{Initial sample selection}

{The goal of this work is to find spatially-resolved sources with analogous colors to the LRDs (e.g. \citealt{Labbe_23a,Labbe_23b,Akins_23,Kokorev_24,PerezGonzalez_24}). Therefore, }the initial sample selection consists of a color cut to collect all the reddest sources in the survey area. We adopt the {same criterion as in \citet{Barro_23} and analogous to that employed by \citet{Akins_23}}:
\begin{equation}
    F277W-F444W>1.5 \, {\rm mag}
\end{equation}
To reduce the possible contamination by fake sources and focus only on the robustly detected galaxies, we couple the color cut with three additional criteria:

\begin{equation}
    F444W< 27.5 \, {\rm mag}   
\end{equation}
\begin{equation}
    S/N(F277W)>2    
\end{equation}
\begin{equation}
    S/N(F444W)>4
\end{equation}
We do not include any S/N requirement in F115W and F150W, since we expect most of the high-\textit{z} or highly dust-obscured sources to dropout at these wavelengths. {Moreover, the depth of COSMOS-Web in the short-wavelength (SW) filters is slightly shallower those of other surveys {where LRDs have been detected} (see e.g. \citealt{Bezanson_22,Finkelstein_23}). Therefore, we do not expect significant emission in these filters even with a similar SED. {Moreover, as noted by \citet{Barro_23}, the inclusion of the additional constrain on the blue colors of LRDs seems to bias the sample towards high-\textit{z} objects. Without this criterion, we aim to focus on the full redshift distribution of these sources. In addition, \citet{Akins_24} -- also selecting (unresolved) LRDs in the COSMOS-Web survey with the same color cut employed here -- showed how this selection can produce a sample of galaxies with properties analogous to those obtained by adding constraints on the blue shape of the SED.} All these criteria produce an initial sample of $\sim450$ sources, after a first vetting to remove imaging artifacts}.

This sample includes a large sample of spatially-compact LRDs \citep{Akins_24}. To focus only on the resolved objects, we perform an initial profile fitting on the F444W maps (i.e., those in which we expect our galaxies to have the highest S/N), forcing the modeling with an unresolved point source (in doing so, we employ the empirical PSF computed on the scientific mosaics through the software \texttt{PSFEx} and the standard routines included in the \texttt{Astropy} library; \citealt{Astropy_22,Bertin_11}). We exclude all galaxies that are well-described as an unresolved point source {(i.e. with a reduced $\chi^2\sim1$ and no significant residual once subtracted the unresolved model)}. The final sample of galaxies consists of 61 objects. The whole sample -- including new objects found in the full 0.54 deg$^2$ -- will be described in a forthcoming paper (Gentile et al., in prep.), and here we focus on the two unique sources among the 61, selected for being particularly high redshift with high stellar mass {estimates} (both of which are described in detail below). Both galaxies, which we named ERD-1 and ERD-2 (for ``extended red dots"), have $F277W-F444W\sim1.75$, just slightly lower than the color cut by \citet{Akins_23}, {but still significantly redder than the nominal LRD selection {(with $F277W-F444W>1$ mag; see e.g. \citealt{Labbe_23a,Greene_23,PerezGonzalez_24})}}. 

\begin{figure*}
    \centering
    \includegraphics[width=0.49\textwidth]{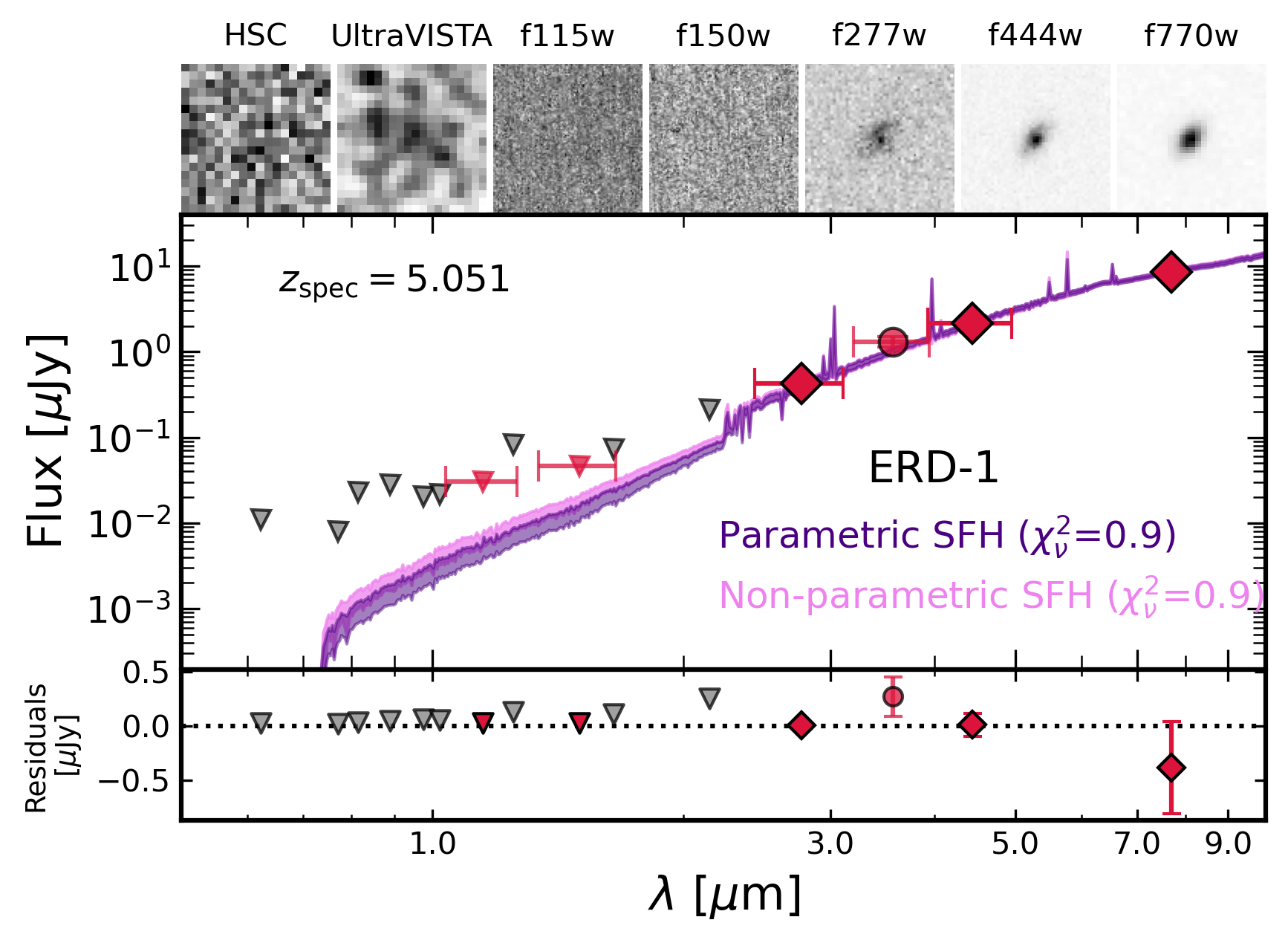}\,\includegraphics[width=0.49\textwidth]{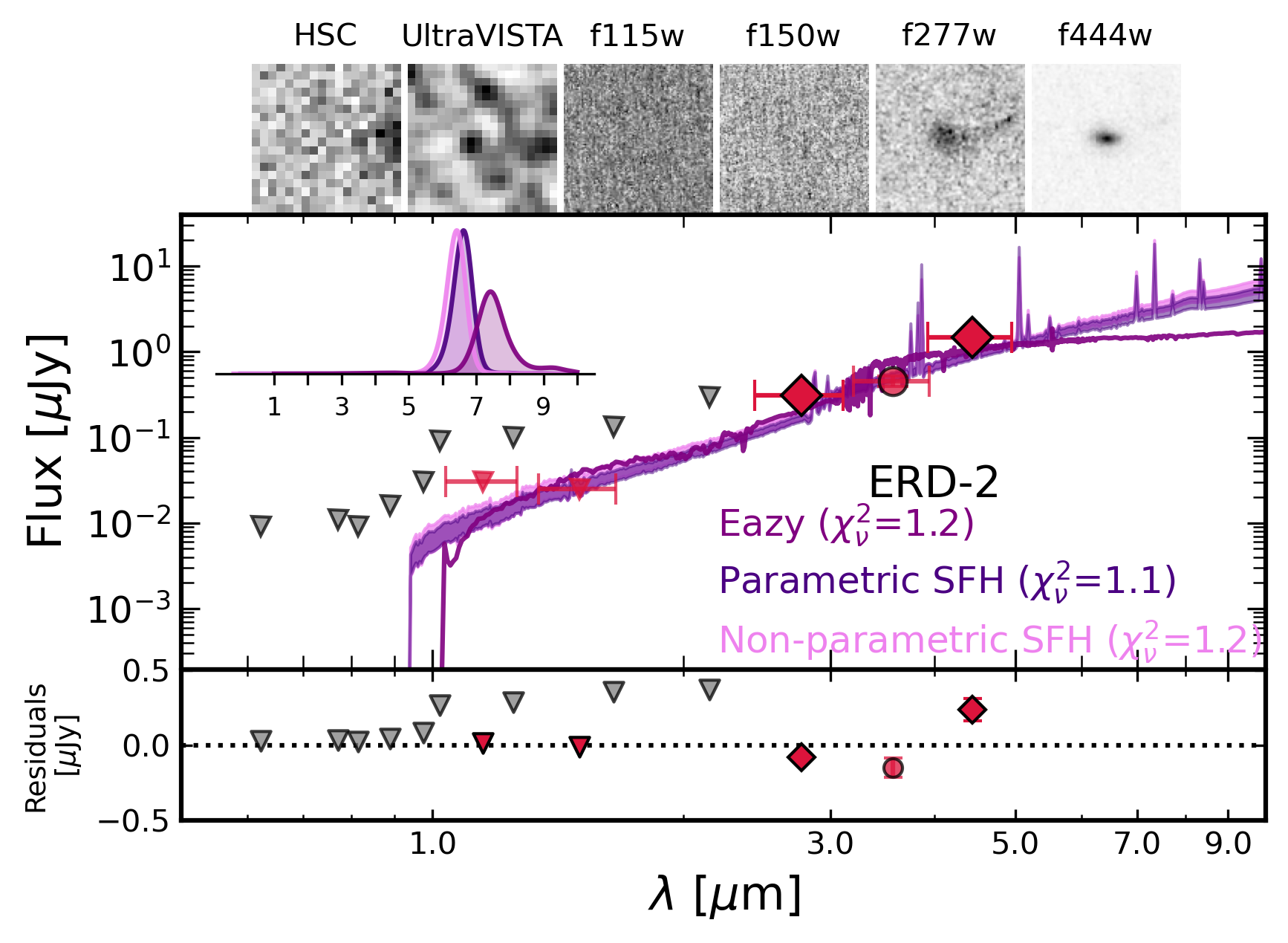}
    \caption{Cutouts (3'' $\times$ 3'') and best fit spectral energy distributions for ERD-1 and ERD-2. Points have $S/N>3$, while downward-pointing triangles represent $2\sigma$ upper limits. {Photometry from JWST is in red, while other constrains are shown in gray.} The first two cutout panels represent stacked images of the $g,r,i,z,y$ bands of the HyperSupremeCam and the $Y,J,H,Ks$ filters of the VISTA telescope, respectively. The inset on the right panel shows the redshift probability distributions for ERD-2 estimated by \texttt{Bagpipes} {(with both SFHs)} and \texttt{Eazy}. Shaded regions mark 16th-84th percentile confidence intervals on SED fits. {The lower panels report the normalized residuals computed on the \texttt{Bagpipes} SED with parametric SFH.}}
    \label{fig:sed}
\end{figure*}

\subsection{Ancillary data}
\label{sec:ancillary}
We extract additional photometry for our sources, using the wealth of other ground/space imaging in COSMOS. We extract fluxes in fixed circular apertures from the datasets in \citet{Weaver_22}. These include the Subaru data (HyperSuprimeCam Subaru Strategic Program, HSC SSP DR3; \citealt{Aihara_19}), VISTA data (UltraVISTA survey DR5; \citealt{McCracken_12}), \textit{Spitzer Space Telescope} data (Cosmic Dawn Survey; \citealt{Moneti_22}), and HST data \citep{Koekemoer_07}. We use a fixed aperture radius of 1'', well-matched to seeing limits of the ground based-data. For the \textit{Spitzer} data, we only include the photometric point at 3.6 $\mu$m as the second channel is superseded by F444W. {We calculate aperture corrections by convolving the best-fitting model by \texttt{SE++} with the IRAC PSF and calculating what fraction of the total flux is missed.}

\subsection{ALMA data on ERD-1}
\label{sec:longwave}

 ERD-1 has a reported spectroscopic redshift $z_{\rm spec}=5.051$ in the literature from \citet{Jin_19} thanks to an ALMA spectral scan. They find a single line at 95.2 GHz ($5.97\sigma$). The \textit{lack} of other lines identified in the broad frequency range lead to the conclusion the line is CO(5-4). As noted in \Sec\ref{sec:sed_fitting}, our new JWST photometry is inconsistent with lower redshift explanations (e.g. CO(4-3) at $z=3.84$ or CO(3-2) at $z=2.63$). \citet{Jin_19} measure a (sub)mm flux density at 3 mm of $S_{\rm 3 mm}=(0.115\pm0.008)$ mJy after removing the spectral line. ERD-1 is also detected at $\lambda=2$ mm in the Ex-MORA survey (Long et al., in prep.), the extended version of the original MORA survey \citep{Casey_21,Zavala_21}, with $S_{\rm 2 mm}=(0.48\pm0.13)$ mJy. We also fit the (sub)mm and FIR photometry (including \textit{Herschel}; \citealt{Liutz_11,Oliver_12}) with the \texttt{mmpz} code \citep{Casey_20}, determining that the probability of a redshift solution at $z<5$ is lower than $9\%$, consistent with the spec-\textit{z} measurement.}

 \subsection{Radio data for ERD-1}
 \label{sec:radio}
 {A radio detection at 3GHz is available for ERD-1 from the VLA-COSMOS large program \citep{Smolcic_17}, with $S_{\rm 3 GHz}=(11\pm2)$ $\mu$Jy. ERD is not detected in the 1.4 GHz maps \citep{Schinnerer_07}, with $S_{\rm 1.4 GHz}=(25\pm19)$ $\mu$Jy or a $2\sigma$ limit of $<63$ $\mu$Jy. This limit implies a radio slope $S\sim\nu^{\alpha}$ with $\alpha>-1.1$}.

\section{SED-fitting}
\label{sec:sed_fitting}

 We derive photometric redshifts with \texttt{Eazy} \citep{Brammer_08} for both galaxies, using the standard set of templates \texttt{tweak\_fsps\_QSF\_12\_v3} \citep{Conroy_10} coupled with those generated by \citet{Larson_23}, and by allowing the redshift to vary in the range [0,10]. We independently estimate the photo-\textit{z} and physical properties with \texttt{Bagpipes} \citep{Carnall_18} using the stellar models from \citet{Bruzual_03}, {leaving the metallicity as a free parameter in the range $[0.1,1]Z_\odot$,} invoking a delayed exponentially declining star formation history (SFH) with a uniform prior on the e-folding time and on the total stellar mass formed in the ranges [0.3,10] Gyr and [$10^5$,$10^{13}$] M$_\odot$. Following \citet{Franco_23}, we use a slightly different implementation of the SFH contained by default in \texttt{Bagpipes}, parametrizing the age of the main stellar population as a function of the Hubble time at a given redshift. We also add a recent instantaneous burst of star formation to the ``secular" SFH taking place in the last $t$ Myr, with a uniform prior on $t$ and on the total stellar mass formed in the burst in the ranges [10,100] and [$0$,$10^{13}$] M$_\odot$. {The synthetic SEDs by \texttt{Bagpipes} also include the emission lines modeled by \texttt{Cloudy} \citep{Ferland_17} assuming a fixed ionization parameter of $\log({\rm U})=-2$}. {To account for the possible dust emission that could be visible in the MIRI filter at high redshift, we also include the models by \citet{Draine_07} with uniform priors on the mass fraction of the Polycyclic Aromatic Hydrocarbons ($q_{\rm PAH}\in[1,4]$) and on the lower limit of starlight intensity distribution ($U_{\rm min}\in[0.1,25]$). The $\gamma$ factor is fixed at the value $\gamma=0.02$}. Finally, SEDs are dust-extincted assuming a \citet{Calzetti_00} law with a uniform prior on $A_{\rm v}$ in the range [0,5] mag and a flat redshift prior from [0,10].

\begin{figure*}
    \centering
    \includegraphics[width=0.45\textwidth]{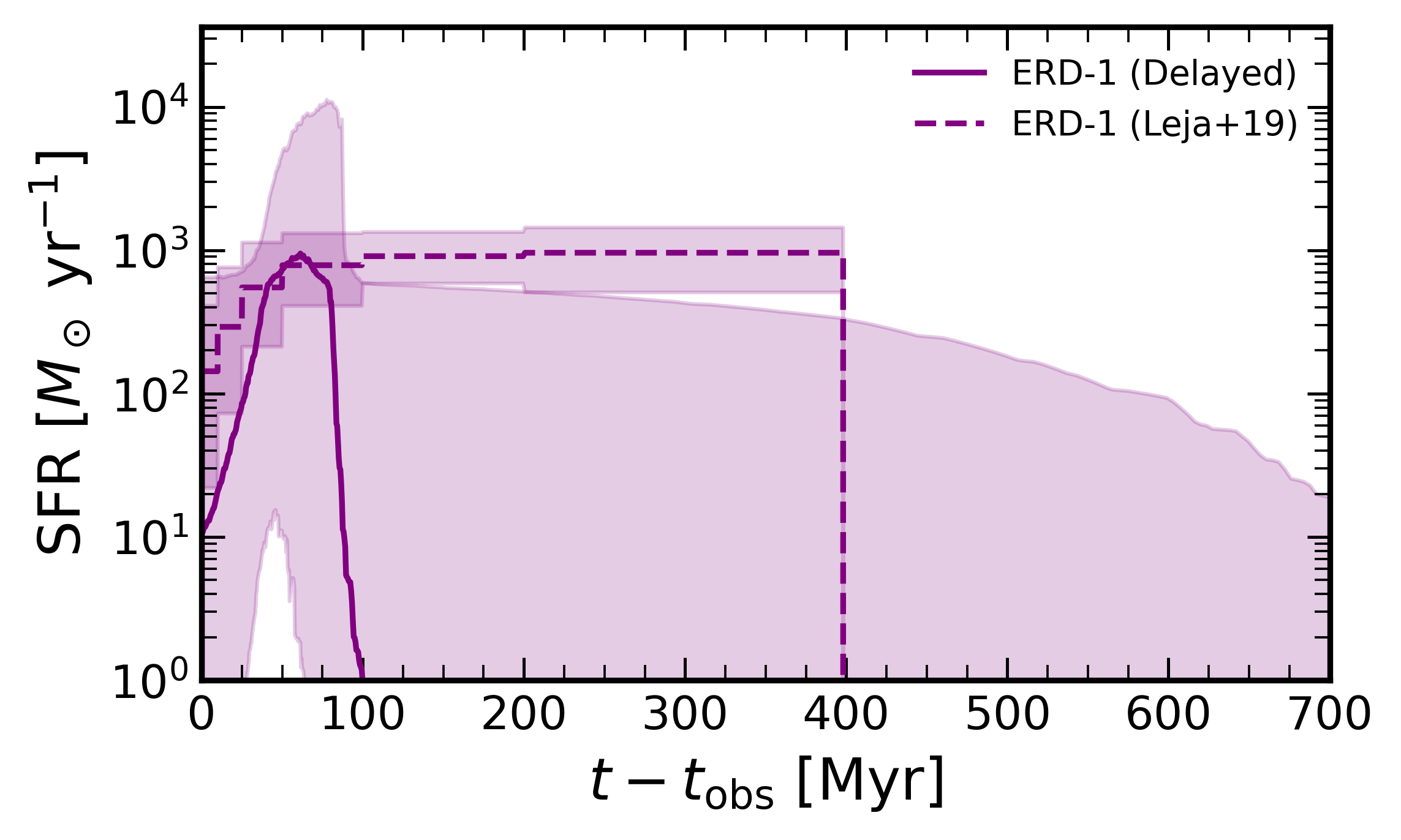}\qquad\includegraphics[width=0.45\textwidth]{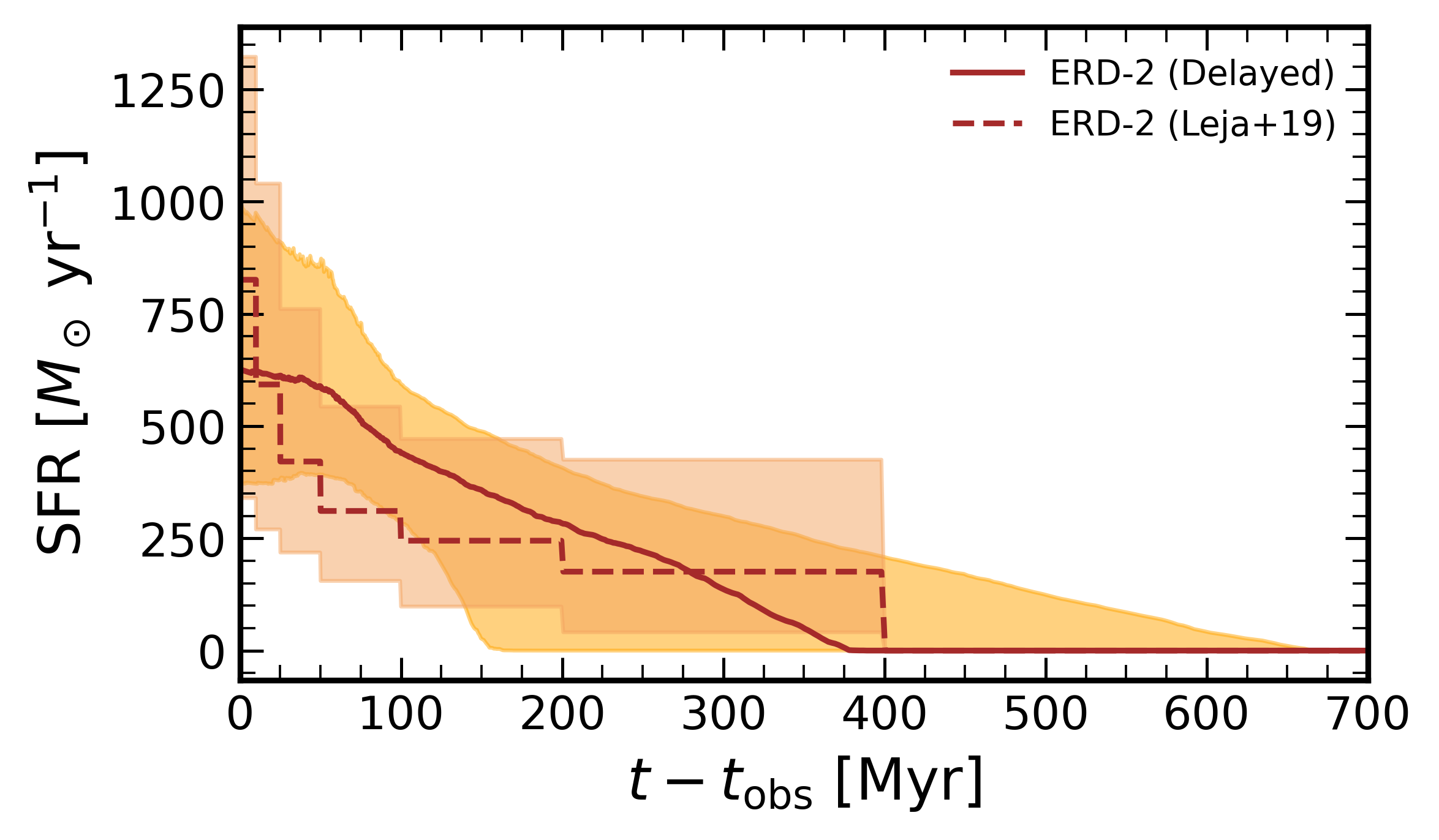}
    \caption{Star formation histories computed for our targets with \texttt{Bagpipes} by assuming a standard delayed exponentially declining SFH (solid lines) and the non-parametric one by \citet{Leja_19}. In both panels, the shaded area reports the 16th-84th percentile of the posterior distribution.}
    \label{fig:SFHs}
\end{figure*}

As a final test, we explore the possible existence of biases due to the chosen parametrization of the star formation history (SFH) of \texttt{Bagpipes} {invoking a non-parametric SFH with the continuity prior as prescribed by \citet{Leja_19}. We compute the $\Delta\log(SFR)$ in six age bins (0-10 Myr, 10-25 Myr, 25-50 Myr, 50-100 Myr, 100-200 Myr, and 200-400 Myr) adopting a prior shaped as a $t-$distribution with $\sigma=0.3$ and $\nu=2$ degrees of freedom. All the other priors are the same as for the parametric SFH.} The output is summarized in \Tab\ref{tab:candidates}, while the best-fitting SEDs are shown in \Fig\ref{fig:sed}. {The inferred SFHs can be found in \Fig\ref{fig:SFHs}}.

\begin{deluxetable*}{ccccccccccc}
\label{tab:candidates}
\tablecaption{Properties of our galaxies estimated through SED-fitting. The table reports, for each objects, the two photometric redshifts computed by \texttt{Bagpipes} ($z_{\rm BG}$) and \texttt{Eazy} ($z_{\rm EZ}$). The stellar mass ($M_\star$) and dust attenuation ($A_{\rm v}$) are estimated with \texttt{Bagpipes} {with both a parametric (P) and non-parametric (NP) SFH.}}
\tablewidth{0pt}
\tablehead{
\colhead{Name} & \colhead{RA} & \colhead{DEC} & \colhead{$z_{\rm spec}$} & \colhead{$z^{\rm P}_{\rm BG}$} & \colhead{$z^{\rm NP}_{\rm BG}$} & \colhead{$z_{\rm EZ}$} & \colhead{$\log(M^{\rm P}_\star)$} & \colhead{$A^{\rm P}_{\rm v}$} &  \colhead{$\log(M^{\rm NP}_\star)$} & \colhead{$A^{\rm NP}_{\rm v}$} \\
\colhead{} & \colhead{(J2000)} & \colhead{(J2000)} & \colhead{$-$}& \colhead{$-$}& \colhead{$-$} & \colhead{$-$} & \colhead{$[M_\odot$]} & \colhead{[mag]} & \colhead{$[M_\odot$]} & \colhead{[mag]}
}
\startdata
ERD-1 & 10:00:47.088 & +02:10:16.680 & $5.051$ & $5.4^{+0.5}_{-0.4}$ & $5.3^{+0.4}_{-0.4}$ & $5.3^{+0.4}_{-0.4}$ & $11.32^{+0.07}_{-0.15}$ & $3.8^{+0.1}_{-0.2}$ & $11.42^{+0.05}_{-0.07}$ & $3.8^{+0.2}_{-0.2}$\\
ERD-2 & 10:01:07.368 & +01:52:00.840 & $-$ & $6.7^{+0.3}_{-0.3}$ & $6.6^{+0.2}_{-0.2}$ & $7.3^{+0.3}_{-0.3}$ & $11.2^{+0.1}_{-0.2}$ & $2.9^{+0.3}_{-0.3}$ & $11.0^{+0.1}_{-0.2}$ & $3.0^{+0.2}_{-0.4}$ \\
\enddata 
\end{deluxetable*}

\subsection{Estimated physical properties}
\label{sec:properties}

{Seen on \Fig\ref{fig:sed}, the two galaxies have similar best-fit SEDs, characterized by a red continuum with a Balmer break and strong emission lines. Indeed, a red continuum with Balmer break explains the high stellar masses and dust extinction (because Balmer breaks imply older stellar populations that have a higher mass-to-light ratio). Moreover, the presence of strong emission lines suggests that these sources are actively forming stars. In addition, the presence of the Balmer break helps secure the photometric redshifts of our sources: the break between the Ks band and the F277W filter (for ERD-1) and between F277W and IRAC channel 1 (for ERD-2) produces the two photometric redshifts of $z\sim5$ and $z\sim6.5$ for the two galaxies, respectively.} 

As visible from \Tab\ref{tab:candidates}, the photo-\textit{z} estimated with \texttt{Eazy} and \texttt{Bagpipes} are in good agreement with each other {(and, for ERD-1, just slightly higher than the spec-\textit{z} measured by \citealt{Jin_19}, albeit compatible within the uncertainties)}. {Since it is known, we fix the redshift of ERD-1 to the spectroscopic redshift.} {Moreover, the physical properties estimated with the non-parametric SFH are in good agreement with those computed by assuming the more standard parametric SFH. In the following, we will assume the latter as reference values for both galaxies.}

{For ERD-1, we find a stellar mass of $\log(M_\star)=(11.32^{+0.07}_{-0.15})$ $M_\odot$ and a dust attenuation of $A_{\rm v}=(3.8^{+0.1}_{-0.2})$ mag. ERD-2 has $\log(M_\star)=(11.2^{+0.1}_{-0.2})$ $M_\odot$ and $A_{\rm v}=(2.9^{+0.3}_{-0.3})$ mag}. The mass for ERD-1 is more accurate than that of ERD-2 thanks to the spec-\textit{z} and the MIRI/F770W detection (tracing the rest-frame NIR wavelengths and -- therefore -- the bulk of the stellar mass, less affected by the presence of dust; see e.g. \citealt{Papovich_23,Wang_24}). {The availability of a MIRI constraint for ERD-2 could, in principle, decrease the inferred stellar mass. \citet{Akins_24} measures offset as high as $\sim$ 1 dex for LRDs with and without the additional MIRI point. However, this effect is primarily due to an overestimated age for the main stellar population \citep{Akins_24}. Since we find a young stellar population for ERD-2 ($\lesssim 400$ Myr; see \Fig\ref{fig:SFHs}), we do not expect this effect to dramatically change our results. To better quantify this possibility, we run an additional test by artificially adding a MIRI/F770W point to the ERD-2 photometry and assuming a ratio [F770W]/[F444W] in the range [1,5], broadly consistent with measured LRD colors at similar redshifts (for instance, the same value for ERD-1 is $\sim$4). By fitting again the new photometry, we obtain a decrease of $\sim$0.4 dex in the inferred stellar mass with a completely flat SED. The offset rapidly decreases with higher F770W fluxes and becomes negligible with [F770W]/[F444W]$\sim$3. This result suggests that the $0.4$ dex decrease should be interpreted as the maximum effect of the additional MIRI constraint on the stellar mass of ERD-2. Nevertheless, a MIRI follow-up of this galaxy would be crucial to properly estimate this value}. {Finally, following \citet{Daddi_10}, we estimate the dynamical mass of ERD-1 from the FWHM of the CO line ($\sim850$ km s$^{-1}$; \citealt{Jin_19}) and the physical size computed in \Sec\ref{sec:morphology}). The obtained value of $\log(M_{\rm Dyn})=(11.3\pm0.2)$ $M_\odot$ is in good agreement with the inferred stellar mass.}

Since neither galaxy is detected in F115W, we have a weak limit on the rest-frame UV flux. Hence, {the values reported by the SED-fitting codes ($\log(SFR)=3.4^{+0.1}_{-0.6}$ $M_\sun$ yr$^{-1}$ and $\log(SFR)=2.8^{+0.2}_{-0.2}$ $M_\sun$ yr$^{-1}$ for ERD-1 and ERD-2, respectively) are largely unconstrained}. However, since ERD-1 has multiple (sub)mm and radio detections -- qualifying it for classic DSFGs selection (see e.g \citealt{Casey_14}) -- and a spectroscopic redshift reducing the possible degeneracies, we estimate the obscured SFR. More specifically, by assuming a standard radio slope of $\alpha=-0.7$ (star-forming galaxies and in good agreement with the upper limit given in \Sec\ref{sec:radio}), we estimate a radio luminosity of $L_{\rm 1.4 GHz}=(3.1\pm0.6)\times10^{24}$ W Hz$^{-1}$. Assuming that all the radio emission is produced by star formation, we use \citet{Kennicutt_12} to estimate a ${\rm SFR}_{\rm 1.4 GHz}=(1.6\pm0.3)\times 10^3 $ M$_\odot$ yr$^{-1}$. This estimate is similar to that computed via the infrared SED in \citet{Jin_19}: $L_{\rm IR}=(6\pm1)\times10^{12}$ $L_\odot$, implying a ${\rm SFR}_{\rm IR}=(1.0\pm0.3)\times 10^3 $ M$_\odot$ yr$^{-1}$. {We measure the FIR-to-radio luminosity ratio as $q_{\rm IR}=2.3\pm0.2$ \citep[see e.g.][]{Helou_85}, in good agreement with $q_{\rm IR}$ of star-forming galaxies \citep[e.g][]{Yun_01}. The high SFR also explains the presence of the strong nebular emission lines visible in the best-fit SEDs.} {All the above-mentioned properties for ERD-1 suggest -- together with the ALMA detections -- that this galaxy belongs to the high-mass end of the dusty star-forming galaxy population (see e.g. \citealt{Casey_14,Wang_19,Gentile_24}), something that is not always true of galaxies with red JWST colors (see e.g. \citealt{Barrufet_24})}.

\section{Morphological Analysis}
\label{sec:morphology}

\begin{figure*}
    \centering
    \includegraphics[width=\textwidth]{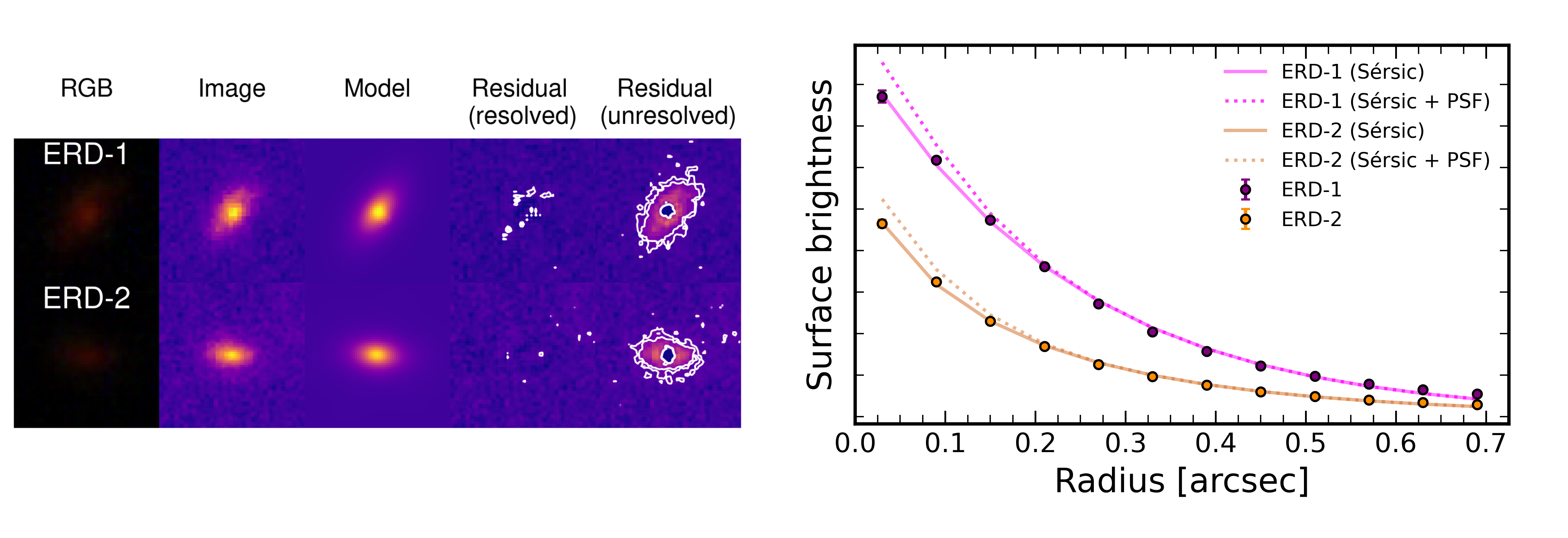}
    \caption{{\textit{Left panel:} RGB cutouts and F444W profile-fitting of ERD-1 and ERD-2 in 2'' $\times$ 2'' cutouts through \texttt{Imfit}. The last two panels show the residuals with the Sérsic model and an unresolved point source. We overplot the contours at 3 and 5 $\sigma$. A single Sérsic profile correctly reproduces the brightness profile of our objects without the need for an additional unresolved component. \textit{Right panel:} Surface brightness profile (in F444W) modeled with a single Sérsic profile and with the addition of an unresolved component. Again, the Sérsic profile well reproduces the observations.}}
    \label{fig:fitting}
\end{figure*}

We use \texttt{Imfit} \citep{Erwin_2015} to perform a two-dimensional profile fitting of our galaxies in F444W, the highest S/N filter, using a single Sérsic model, leaving the centroid, axis ratio, position angle, Sérsic index, effective radius, and flux as free parameters. Our resulting models describe the data well, with reduced $\chi^2\sim1$ (see \Fig\ref{fig:fitting}). {From the profile fitting, we obtain that ERD-1 and ERD-2 have effective radii of $(1.87\pm0.02)$ kpc and $(1.43\pm0.02)$ kpc respectively, and that both are well described by a Sérsic profile with $n\sim1$. }

We also fit the data to a combined model with both an unresolved point source and Sérsic profile. To limit the additional free parameters, we fix the unresolved centroid  to the center of the Sérsic model, and leave the flux as a free parameter. The best-fitting model always converges towards a zero-flux point source. {If we fix the contribution from the point source to be at least $\sim20\%$ of the total integrated flux (see e.g. the right panel in \Fig\ref{fig:fitting}), we obtain a significantly worse fit, clearly indicating no contribution from an unresolved component.}

\section{Discussion}
\label{sec:discussion}

\subsection{Estimating the stellar baryon fraction}
\label{sec:cosmology}

  $\Lambda$CDM cosmology uniquely determines a certain mass function of dark matter halos at a given redshift \citep[e.g.][]{Sheth_99}. However, converting this halo mass to a stellar mass is a more challenging task, since baryonic processes are complex and not completely understood, especially at high-\textit{z}. Under the oversimplified hypothesis that all the baryons in halos are converted in stars (an assumption that is most certainly incorrect), we can derive a stringent upper limit on the galaxies' possible masses at each redshift in a given cosmological volume (see e.g. \citealt{Steinhardt_16,Behroozi_18,Boylan_23}) and compare our estimated stellar masses with these limits.
 
 Here we follow the analysis of \citet{Boylan_23}\footnote{We use the github code referenced therein, provided at this link: \url{https://github.com/mrbk/JWST_MstarDensity/tree/main}}. We first consider the halo mass function by \citet{Sheth_99} and multiply it by the cosmic baryon fraction $f_{\rm b}\sim0.16$ \citep{Planck_20} to obtain a relationship between the maximal stellar mass and volume density: $M_\star=M_{\rm H}f_{\rm b} \epsilon_\star$. This value depends on $\epsilon_\star$, the stellar baryon fraction, describing the integrated history of the star formation efficiency in a given galaxy.
 
\begin{figure}
    \centering
    \includegraphics[width=\columnwidth]{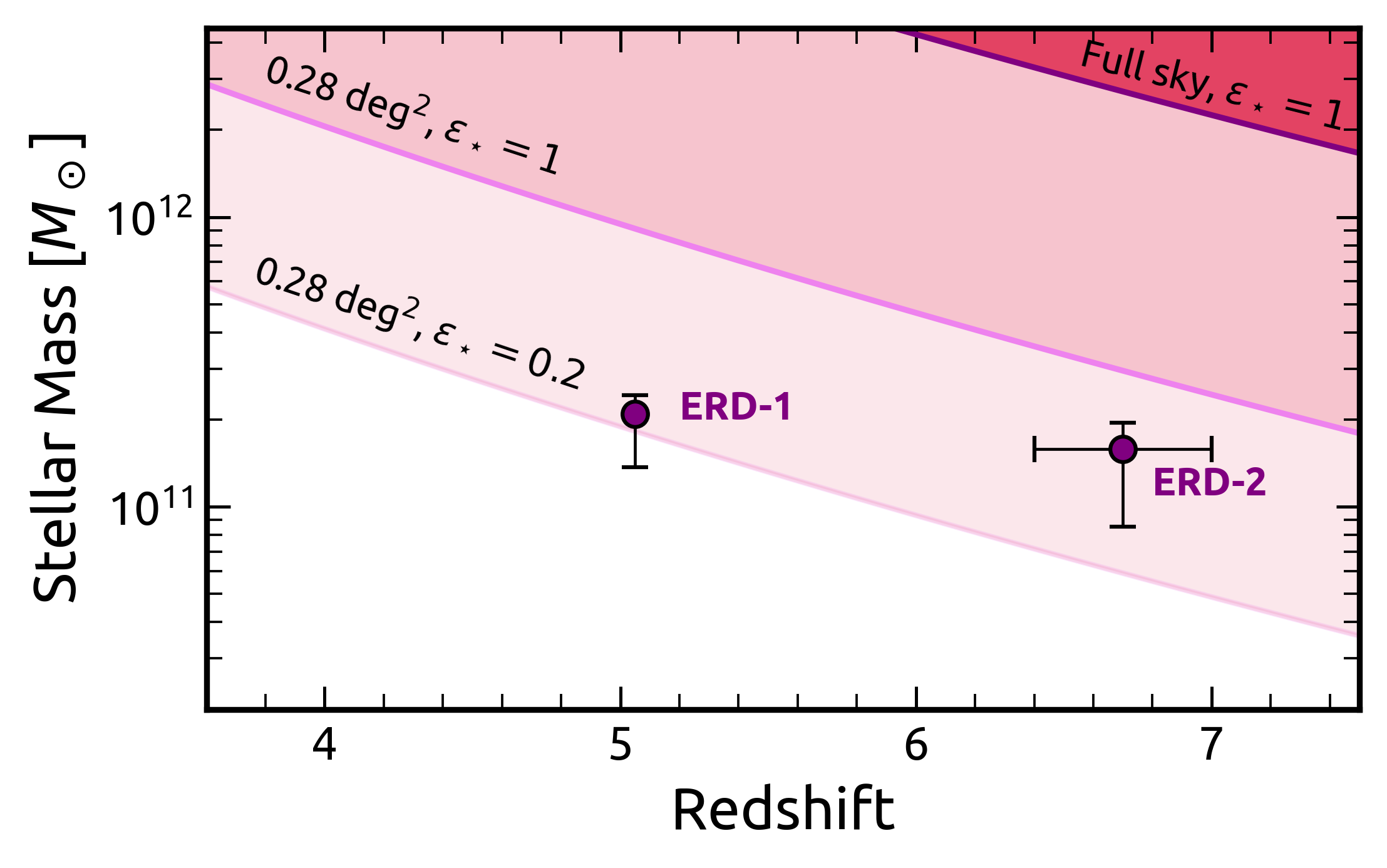}
    \caption{ERD-1 and ERD-2 in the stellar mass vs. redshift plane. The lines correspond to different number densities: $1.3\times10^{-12}$ Mpc$^{-3}$ (i.e. a single galaxy in the whole sky; dark violet) and $1.9\times10^{-7}$ Mpc$^{-3}$ (i.e. one source in the cosmic volume explored so-far by COSMOS-Web; dark red) with a (unrealistic) stellar baryon fraction of $\epsilon_\star=1$. The lower (light pink) line scales down the second using a more reasonable efficiency of $\epsilon_\star=0.2$.}
    \label{fig:cosmology}
\end{figure}

 {Considering the redshift range spanned by our galaxies ($5<z<7$) and the sky area analyzed here from COSMOS-Web (0.28 deg$^2$), we estimate {that our survey covers a total comoving volume} of $7.6\times10^{11}$ Mpc$^3$}. In \Fig\ref{fig:cosmology}, we compare our sources' stellar masses with the maximum mass allowed with redshift in our survey volume. To fit within $\Lambda$CDM, the two galaxies presented in this paper would require a (minimum) {$\epsilon_\star=0.24^{+0.02}_{-0.08}$ and $\epsilon_\star=0.5^{+0.2}_{-0.2}$ (for ERD-1 and ERD-2, respectively)} under the hypothesis that they are the only galaxies in the cosmic volume explored so far by COMSOS-Web with these masses at these redshifts.

\subsection{Implication for star formation efficiency}

 Star formation is known to be very inefficient, with several self-regulating processes limiting the efficiency at small scales (see e.g. \citealt{Kennicutt_98,Ostriker_11}). Observational studies at lower redshifts {($z\lesssim2-3$)} have an implied strong upper limit on $\epsilon_\star\ll0.2$ \citep[e.g.][]{Mandelbaum_06,Behroozi_10,Kennicutt_12,Moster_13,Wechsler_18,Shuntov_22}, with some (e.g. \citealt{Behroozi_10} and \citealt{Moster_13}) notably expecting a decreased integrated star formation efficiency towards higher redshifts {and others finding a more constant behavior until at least $z\sim4$ (e.g. \citealt{Behroozi_13})}. Our values of $\epsilon_\star$ are significantly higher than these studies expect, suggesting star formation pathways much more efficient at high-\textit{z} than what is observed in the local Universe. {This conclusion is strengthened by similar $\epsilon_\star$ reported by \citet{Xiao_23} for their red galaxy sample (with spectroscopic redshifts) from the FRESCO survey \citep{Oesch_23}, suggesting $\epsilon_\star\sim0.5$, {or with those reported by \citet{Wang_24} for a sample of MIRI-detected massive galaxies, with a moderately lower $\epsilon_\star\sim0.3$.} 

 We underline, however, that the estimated $\epsilon_\star$ are also the product of some assumptions that -- if proved wrong -- could strongly affect the measured values. Firstly, the halo mass function is a consequence of the assumed $\Lambda$CDM cosmology. Some alternative models (e.g. early dark energy; see e.g. \citealt{Karwal_16}, \citealt{Poulin_18}) forecast the early formation of more massive halos \citep[e.g.][]{Klypin_21}. In this case, halos are more massive, and a larger baryonic mass would be available for star formation.

 Secondly, the stellar masses obtained in this paper assume a \citet{Chabrier_03} IMF. Several works argue that this could be an oversimplification, due to the dependence of the IMF on the metallicity or the temperature of molecular clouds (see e.g. \citealt{Low_76,Dave_08}). A top-heavy IMF could produce lower stellar masses {(with a factor $\gtrsim0.3$ dex; see e.g. \citealt{Woodrum_23,Wang_24_IMF})} and give slightly lower values of $\epsilon_\star$. 

 {Moreover, the possible contribution of reddened AGN emission to the continuum can significantly alter the stellar mass estimation (even with a spectroscopically confirmed Balmer break; see e.g. \citealt{Wang_24_rubies}). This hypothesis is ruled out for our galaxies thanks to their extended morphology, ensuring that the main contributor to the red continuum is the stellar emission.}

 {Finally, even excluding the contribution of AGN, the presence of strong emission lines (due to intense star formation) can produce an overestimation of the stellar masses of galaxies (see e.g. \citealt{Schaerer_09}). Unfortunately, the contribution of these lines is hard to constrain with photometry alone. Therefore, a spectroscopic follow-up with JWST would be necessary to further confirm the inferred physical properties of our targets.}

 While some simulations suggest that very efficient star formation may be possible in dense giant molecular clouds (e.g. \citealt{Torrey_17,Grudic_18,Dekel_23}), the extended morphology in these cases seems somewhat disjoint from such an explanation. Further observations of the mass budget in such systems e.g. from spatially resolved observations of [CII] dynamics from ALMA, may be crucial to deciphering the underlying puzzle.

\section{Conclusions}
\label{sec:conclusions}

In this paper, we presented two {candidate} massive, extended and highly dust-obscured galaxies: ERD-1 and ERD-2. These sources have extremely red ($F277W-F444W\sim1.7$ mag) color, mimicking the selection of ``little red dots" \citep[e.g.][]{Akins_23,Barro_23}. However, these galaxies have an extended morphology, ruling out the possibility that the continuum is dominated by reddened nuclear activity (unlike the majority of LRDs) thus ensuring that the stellar emission is responsible for the observed red colors.

{We have employed SED fitting using two different codes (\texttt{Eazy} and \texttt{Bagpipes}) yielding photo-\textit{z} estimates of ERD-2 of $z\sim6.7$, while a spec-\textit{z} of 5.051 of ERD-1 \citep{Jin_19} is already published.} Both the best-fitting SEDs show a red continuum with a Balmer break combined with emission lines. These features suggest a moderately aged stellar population with high stellar masses ($M_\star\sim10^{11}$ M$_\odot$) and dust extinction ($A_{\rm v}>2.5$ mag), together with an ongoing episode of significant star formation.

The high stellar masses, once placed in the cosmological context through comparison with the halo mass function, suggest that our galaxies {could} have a much higher stellar baryon fraction ($\epsilon_\star\gtrsim0.2$) than what is commonly observed in lower-redshift sources.

Possible explanations of this result are a top-heavier IMF in the high-\textit{z} Universe, a much more efficient star formation in the early stages of the cosmic history than has been constrained by observations at low-\textit{z}, or an earlier formation of massive halos than what predicted by $\Lambda$CDM.

These possible consequences increase the need for spectroscopic follow-up of these interesting systems, {to confirm the photo-\textit{z} (for ERD-2), {confirm the presence of the Balmer breaks suggested by the photometry (e.g. \citealt{Wang_24_rubies})}, and to accurately measure the dynamical masses of our galaxies, which is fully possible using JWST directly or ALMA.}

\begin{acknowledgments}
 We thank the anonymous referee for their comments that improved the initial version of this study. F.G. thanks the department of astronomy of the University of Texas at Austin for the hospitality during the writing of this paper. F.G. and C.M.C. thank Zhaoxuan Liu and Seiji Fujimoto for the careful reading of a first draft of this letter. F.G. and M.T. acknowledge the support from grant PRIN MIUR 2017-20173ML3WW\_001. ‘Opening the ALMA window on the cosmic evolution of gas, stars, and supermassive black holes’. Support for this work was provided by NASA through grant JWST-GO-01727 awarded by the Space Telescope Science Institute, which is operated by the Association of Universities for Research in Astronomy, Inc., under NASA contract NAS5-26555. C.M.C., H.A., M.F., J.M., and A.S.L. acknowledge support from the National Science Foundation through grants AST-2307006, AST-2009577, and the UT Austin College of Natural Sciences for support. C.M.C. also acknowledges support from the Research Corporation for Science Advancement from a 2019 Cottrell Scholar Award sponsored by IF/THEN, an initiative of the Lyda Hill Philanthropies. {J.R. was supported by the Jet Propulsion Laboratory, California Institute of Technology, under a contract with the National Aeronautics and Space Administration (80NM0018D0004). D.L. acknowledges the support from the Strategic Priority Research Program of the Chinese Academy of Sciences, grant No. XDB0800401.} This work used the CANDIDE computer system at the IAP supported by grants from the PNCG, CNES, and the DIM-ACAV and maintained by S. Rouberol. {The JWST data presented in this article were obtained from the Mikulski Archive for Space Telescopes (MAST) at the Space Telescope Science Institute. The specific observations analyzed can be accessed via \dataset[doi: doi:10.17909/ahg3-e826]{https://doi.org/10.17909/ahg3-e826}.}
\end{acknowledgments}

\bibliography{sample631}{}
\bibliographystyle{aasjournal}

\end{document}